\documentclass[11pt,a4paper]{article}

\usepackage{euler,beton}
\usepackage{amsmath,amsthm,amssymb}
\usepackage{xspace}

\newcommand{\Nat}{\mathbb{N}}                    
\newcommand{\monus}{\dot-}
\def\vek#1{\overrightarrow{#1\,\,}}

\newcommand{\var}[1]{\alpha_{#1}}                
\newcommand{\typ}[1]{\tau_{#1}}                  
\newcommand{\typp}{\tau'_0}                      
\newcommand{\fv}[1]{\mathsf{FV}({#1})}           
\newcommand{\abs}[2]{\lambda {#1}\, .\, {#2}}    
\newcommand{\pair}[2]{\langle {#1},{#2}\rangle}  
\newcommand{\pl}[1]{{#1}\mathsf{L}}              
\newcommand{\pr}[1]{{#1}\mathsf{R}}              
\newcommand{\conv}{\mathrel{\mapsto}}            
\newcommand{\rd}{\mathrel{\leadsto}}             
\newcommand{\all}[2]{\forall {#1} . {#2}}        
\newcommand{\subst}[3]{{#1}[{#2}:={#3}]}         
\newcommand{\ass}[2]{{#1}\mathbin{:}{#2}}        
\newcommand{\judge}[3]{{#1}\mathrel{\vdash}\ass{#2}{#3}}  

\newcommand{\ax}{\ensuremath{(\text{var})}\xspace} 
\newcommand{\impi}{\ensuremath{(\to\! I)}\xspace}  
\newcommand{\impe}{\ensuremath{(\to\! E)}\xspace}  
\newcommand{\prodi}{\ensuremath{(\times I)}\xspace}  
\newcommand{\prodel}{\ensuremath{(\times E^1)}\xspace}  
\newcommand{\proder}{\ensuremath{(\times E^2)}\xspace}  
\newcommand{\alli}{\ensuremath{(\forall I)}\xspace}  
\newcommand{\alled}{\ensuremath{(\forall E^1)}\xspace}  
\newcommand{\alles}{\ensuremath{(\forall E^2)}\xspace}  

\newcommand{\nat}{\mathsf{Nat}}                  
\newcommand{\natz}{\nat_0}
\newcommand{\nati}{\nat_1}
\newcommand{\typn}[1]{{#1}^{\ast}}               
\newcommand{\typnn}[1]{{#1}^{\ast\ast}}          
\newcommand{\typi}[2]{{#1}^{({#2})}}             
\newcommand{\num}[1]{\underline{#1}}             

\newcommand{\suc}{\mathop{\mathsf{suc}}}        
\newcommand{\add}{\mathop{\mathsf{add}}}        
\newcommand{\mul}{\mathop{\mathsf{mult}}}       
\newcommand{\pred}{\mathop{\mathsf{pred}}}      
\newcommand{\cd}{\mathop\downarrow\nolimits}    
\newcommand{\cu}{\mathop\uparrow\nolimits}      
\newcommand{\sub}{\mathop{\mathsf{sub}}}    
\newcommand{\subt}{\mathop{\widetilde{\mathsf{sub}}}\nolimits}   

\newcommand{\rk}[1]{\mathrm{rk}({#1})}         
\newcommand{\ajudge}[5]{{#1}\mathrel{\vdash^{#2}_{#3}}\ass{#4}{#5}}  

\newcommand\betaeq{\mathrel{{=}_\beta}}
\newcommand\betaetaeq{\mathrel{{=}_{\beta\eta}}}

\newtheorem{theorem}{Theorem}
\newtheorem{lemma}[theorem]{Lemma}
\newtheorem{proposition}[theorem]{Proposition}
\newtheorem{corollary}[theorem]{Corollary}
\newtheorem*{definition}{Definition}

\begin{document}

\title{An Elementary Fragment of Second-Order Lambda Calculus}
\author{Klaus Aehlig\thanks{Supported by the DFG Graduiertenkolleg
    ``Logik in der Informatik''}\\ Mathematisches Institut\\ Universit\"at
  M\"unchen \and Jan Johannsen\thanks{Partially supported by DFG grant
    Jo 291/2-2}\\ Institut f\"ur Informatik\\
  Universit\"at M\"unchen}
\date{~}

\maketitle 

\allowdisplaybreaks

\begin{abstract}
  A fragment of second-order lambda calculus (System $F$) is defined
  that characterizes the elementary recursive functions.  Type
  quantification is restricted to be non-interleaved and stratified,
  i.e., the types are assigned levels, and a quantified variable can
  only be instantiated by a type of smaller level, with a slightly
  liberalized treatment of the level zero.
\end{abstract}
  
\begin{section}{Introduction and Related Work}

Machine-independent characterizations of computational complexity
classes are at the core of the research area called \emph{Implicit
Computational Complexity} which has received a lot of attention
recently. The goal is to give natural descriptions of these classes
by \emph{conceptual} means, mostly derived from mathematical logic. In
particular it is desirable to go without any explicit mention of
bounds or ad hoc initial functions.

The second-order, or polymorphic lambda calculus (System $F$) 
\cite{gir71,rey74}
provides a particularly natural framework for this purpose, as all
data-types, such as natural numbers, binary words or trees, can be
encoded therein without the use of constructors or initial functions.
Unfortunately, full system $F$ has a computational strength far beyond
anything reasonable in this context: all functions provably total in
second-order arithmetic can be defined.

Recently there have been approaches to define fragments of system $F$
with a weaker
computational strength. Altenkirch and Coquand~\cite{altcoq01}
proposed a fragment characterizing the functions provably recursive in
Peano Arithmetic by restricting type abstraction to first-order types
in a single variable.  Earlier, Leivant~\cite{leiv91} has used
stratification of type abstraction to obtain a fragment characterizing
the fourth level $\mathfrak E_4$ of the Grzegorczyk
hierarchy~\cite{grz53}.

Here we give a characterization of the third level $\mathfrak E_3$ of
the Grzegorczyk hierarchy, that is, the Kalm\'ar elementary recursive
functions. In order to achieve this, we use a stratification of type
abstractions into only two levels. This alone would give a system in
which all definable functions are elementary recursive. However, the
class would presumably not be exhausted, as, for example, subtraction
seems to be undefinable.

Therefore we use a primitive product type former
and allow a quantified variable of the lowest level to be
instantiated by a finite product of itself. Note that product types
are definable in system $F$, however using an additional type
quantifier and thus disturbing our stratification.

Different restrictions of system $F$ based on linear logic, and
characterizing also the elementary recursive functions, as well as
polynomial time, were introduced by Girard~\cite{gir98} and further
elaborated by several authors~\cite{asprov02,danjoi02}.
\end{section}

\begin{section}{Definitions}

The elementary recursive functions are a natural subclass of the
primitive recursive functions that was first defined by Kalm\'ar
\cite{kalmar}.  
A function $f(x,\vec{y})$ is a bounded sum (a bounded product), if it
is defined from $g(x,\vec{y})$ by 
\[ f(x,\vec{y}) =\sum_{i=0}^{x-1} g(i,\vec{y})  \qquad \Bigl(\text{
  resp. } \quad f(x,\vec{y}) =\prod_{i=0}^{x-1} g(i,\vec{y})
\;\;\Bigr) \; . \] The elementary recursive functions are the least
class of number-theoretic functions that contains the constant $0$,
all projections, successor, addition, modified subtraction $x\monus y
:= \max(x-y,0)$, multiplication and is closed under composition and
bounded sums and products.

It is well-known that the elementary recursive functions coincide with
the third level $\mathfrak E_3$ of the Grzegorczyk hierarchy
\cite{grz53}, and that they coincide as well with the functions
computable in time or space bounded by an elementary recursive
function (see e.g.~\cite{clotehb}). 

The functions $\abs{n}{2_k(n)}$ for $k \in \Nat$ are inductively
defined as follows: $2_0(n) = n$ and $2_{k+1}(n) = 2^{2_k(n)}$\@.  
For every fixed $k$, this function is elementary recursive, but the binary
function $\abs{kn}{2_k(n)}$ is not: 
$\abs k {2_k(1)}$ eventually majorizes every elementary recursive
function.

\paragraph{The system.} 
We now give a formal definition of our system, by means of a type
assignment calculus. So terms are only the terms of the untyped lambda
calculus with pairs, i.e., given by the grammar
$$r,s ::= x \mid rs \mid \abs{x}{r} \mid \pair{r}{s} \mid \pl{r}
  \mid \pr{r}\; , $$
where $x$ ranges over an infinite set of variables.
We define types of
level $n$ for a natural number $n$\@. However, we will use only the
types of level at most $2$\@. Our type variables also come in different
levels; let $\var n$ range over variables of level $n$.
\begin{definition}
  The types $\typ n$ of level $n$ and the flat types $\typp$ of
  level $0$ are inductively given by the following grammar:
\begin{align*}
  \typ{n} & := \var{n} \mid \typ{n} \to \typ{n} \mid \typ{n} \times \typ{n} \mid \all{\var{k}}{\typ{k}} \\
  \typp &:= \var{0} \mid \typp \times \typp 
\end{align*}
where $k<n$ and $\fv{\typ{k}} \subseteq \{ \var k \}$. 
\end{definition}
Note that this notion of the level of a type differs from the notion
commonly used in the literature, so it should more correctly be called
\emph{modified level}. However, since the usual notion is not used in
the present work, for sake of brevity we just use the term
\emph{level} for the modified notion.

Also note that with respect to our notion of the level of a type, there
are no closed types of level $0$. 

\paragraph{Contexts and Judgments.}
A context $\Gamma$ is a set of pairs $x:\tau$ of variables and types,
where the variables occurring in a context have to be distinct.
A typing judgment is of the form $\judge{\Gamma}{r}{\tau}$ and
expresses that $r$ has type $\tau$ in the context $\Gamma$\@. 
The typing rules are:
\begin{align*}
  \ax\;&\dfrac{}{\judge{\Gamma}{x}{\tau}} & &\text{if }x:\tau\text{ occurs in }
  \Gamma 
\\[2ex] 
  \impi\;&\dfrac{\judge{\Gamma,x:\sigma}{r}{\rho}}{\judge{\Gamma}{\abs
      x r}{\sigma\to\rho}} 
& 
  &\impe\;\dfrac{\judge{\Gamma}{r}{\sigma \to \rho} \qquad
    \judge{\Gamma}{s}{\sigma}}{\judge{\Gamma}{rs}{\rho}} 
\\[2ex] 
  \prodi\;&\dfrac{\judge{\Gamma}{r}{\rho} \qquad \judge\Gamma s \sigma}{\judge\Gamma{\pair{r}{s}}{\rho\times\sigma}} 
\\[2ex] 
  \prodel\;&\dfrac{\judge\Gamma r {\sigma\times\rho}}{\judge\Gamma{\pl r}\sigma} 
 &  
  &\proder\;\dfrac{\judge\Gamma r {\sigma\times\rho}}{\judge\Gamma{\pr r}\rho} 
\\[2ex]
  \alli\;&\dfrac{\judge{\Gamma}{r}{\typ k}}{\judge{\Gamma}{r}{\all{\var k}{\typ k}}} 
  & &\text{if } \var{k} \notin \fv{\Gamma}
\\[2ex]  
  \alled\;&\dfrac{\judge{\Gamma}{r}{\all{\var k}{\typ k}}}{\judge{\Gamma}{r}
  {\subst{\typ k}{\var k}{\sigma_\ell}}} & 
  &\text{if } \ell\leq k \text{ and } \fv{\sigma_\ell} = \emptyset 
\\[2ex]
  \alles\;&\dfrac{\judge\Gamma r {\all{\var 0}{\typ 0}}} {\judge\Gamma
    r {\subst{\typ 0}{\var 0}{\sigma'_0}}}
  & &\text{where }  \sigma'_0 \text{ is  flat type.} 
 \end{align*}
We will tacitly use the obvious fact that $\judge\Gamma{r}\tau$ holds
only if all the free variables of $r$ are assigned a type in
$\Gamma$.   
The rules are formulated in such a way that weakening is
admissible. By a simple induction on the derivation one verifies:
\begin{proposition}[Weakening] 
  If $\Gamma \subseteq \Gamma'$ and $\judge{\Gamma}{r}{\tau}$, then
  $\judge{\Gamma'}{r}{\tau}$.
\end{proposition}

\paragraph{Reductions.}
Our system is equipped with the usual reductions of lambda calculus
with pairs. Let $\rd$ be the reflexive and transitive closure of the
reduction given by the compatible closure of the conversions below,
i.e., by allowing application of these conversions to arbitrary
subterms.
\begin{align*}
 (\abs{x}{r})s  &\conv \subst{r}{x}{s} \\ 
 \pl{\pair{r}{s}} &\conv r \\
 \pr{\pair{r}{s}} &\conv s \; . 
\end{align*}
We denote the induced congruence relation by $\betaeq$, i.e.,
$\betaeq$ is the symmetric and transitive closure of $\rd$. For
technical reasons, in some proofs we will also need the notion of
$\beta\eta$-equality, denoted by $\betaetaeq$. It is defined like
$\betaeq$, but based on the conversions above together with
$\eta$-conversion  
\[ \abs x {tx}\conv t\]
 with the proviso that $x$ is not free in $t$.

It is easily verified that our reductions preserve typing.
\begin{proposition}[Subject Reduction]
  If $\judge\Gamma r\tau$ and $r\rd r'$, then $\judge\Gamma {r'}\tau$. 
\end{proposition}

\paragraph{Statement of the main result.}
For every type $\tau$, we define the type $$\typn{\tau} := (\tau \to
\tau)\to(\tau\to\tau)\,.$$
For a natural number $n$, the Church
numeral $\num n$ is $\abs{fx}{f^nx}$, it can have type $\typn{\tau}$
for every $\tau$\@.  The types of natural numbers are $\natz :=
\all{\var 0}{\typn{\var 0}}$ and $\nati := \all{\var 1}{\typn{\var
    1}}$\@. It can be shown that the only closed normal inhabitants of
the types $\nat_i$ are the Church numerals, and the identity
combinator $\mathsf{id} := \abs{x}{x}$, which is equivalent to the
numeral $\num 1$ under $\eta$-conversion.

A function $f\colon\Nat^k \to \Nat$ is \emph{representable}, if there
is a term $t_f$ such that \mbox{$\judge{}{t_f}{\nati^k \to \natz}$}
and for all $\vec{n} \in \Nat^k$, it holds that $t_f \vec{\num n}
=_\beta \num{f(\vec n)}$\@. We shall prove below, as Corollary \ref{char},
that the representable functions are exactly the elementary recursive
functions.

\paragraph{Notation.} 
As usual, lists of notations for terms, numbers etc.\ that only differ
in successive indices are denoted by leaving out the indices and
putting an arrow over the notation. It is usually obvious where to add
the missing indices, otherwise we add dots wherever an index is left
out.  We use one dot if the index runs with the innermost arrow, two
dots if the index runs with the next innermost arrow etc., so that
e.g.\ the expression 
\[ \vek{ t_\cdot \vek{n_{\cdot\cdot,\cdot}} } \] 
stands for a sequence of the form 
\[ t_1  n_{1,1}  \ldots  n_{1,k_1} \; , \; \ldots  \; , \; t_r
n_{r,1} \ldots  n_{r,k_r} \; . \] 
\end{section}

\begin{section}{Completeness}

In this section we show one direction of our claim, namely we show
that every elementary recursive function can be represented by a term.
To start, it is easy to check that the usual basic arithmetic
functions can have the following types
$$\begin{array}{lclcl}
  \suc &:=& \abs{nsz}{s(nsz)} &:& \typn{\tau}\to\typn{\tau} \\
  \add &:=& \abs{mnsz}{ms(nsz)} &:& \typn{\tau}\to\typn{\tau}\to\typn{\tau} \\
  \mul &:=& \abs{mns}{m(ns)}  &:& \typn{\tau}\to\typn{\tau}\to\typn{\tau} \\
\end{array}$$
{for every $\tau$\@. We use these to program a downward
  typecast, that is a function }
$$
  \cd\, :=\,\abs{n}{n\suc\num 0} \,\,\colon \typnn{\tau} \to \typn{\tau} 
$$
with the property $\cd \num n\betaeq \num n$\@. Note that $\cd$ 
also has the type $\typn{\natz}\to\natz$, since $\suc$ can be typed as
$\natz\to\natz$ by instantiating the argument $\ass{n}{\natz}$ as
$\ass n{\typn{\var0}}$. Note moreover that $\add$ and $\mul$, by a similar
argument, can also have type $\natz\to\natz\to\natz$.
 
The predecessor can be implemented of type $\natz\to\natz$ as follows:
in the context where we have variables $\ass{s}{\var0\to\var0}$ and
$\ass{z}{\var0}$, as abstract successor and zero, we get the term
$P := \abs{p}{\pair{s(\pl{p})}{\pl{p}}}$ of type
$(\var0\times\var0)\to (\var0\times\var0)$, such that the $n$-fold
iteration of $P$ applied to $\pair zz$ reduces to $\pair{s^n
  z}{s^{n\monus 1}z}$, for every $n\geq 0$. 
Thus the argument $\ass{n}{\natz}$ is
instantiated as $\ass{n}{\typn{(\var0\times\var0)}}$ by the rule
\alles, and we get
\[
\judge{\ass{n}{\natz}}{\abs{sz}{\pr{n P \pair{z}{z}}}}{\typn{\var0}}
\] 
and an application of \alli and \impi yields that the predecessor
$\pred := \abs{nsz}{\pr{n P \pair{z}{z}}}$ is typeable as
$\ass{\pred}{\natz\to\natz}$. 

We obtain subtraction $\sub := \abs{mn}{n\pred m}$ by iterating the 
predecessor, of type $\ass{\sub}{\natz \to \typn\natz \to \natz}$.
Obviously, for $m,n\in \Nat$ we have $\sub \,\num m \,\num n
\betaetaeq \num{m \monus n}$. 

Testing for zero can also be easily programmed as $\chi_0 :=
\abs{nxy}{n(\abs{z}{y})x}$, which has type $\ass{\chi_0}{\natz \to
  \var0\to\var0\to\var0}$, and the operational semantics \textsl{if
  $n=0$ then $x$ else $y$}, i.e., with the properties $\chi_0 \, \num
0 \, x \, y \betaeq x$ and $\chi_0 \, \num{n+1} \, x \, y \betaeq y$.
To obtain the typing, we instantiate the input $\ass{n}{\natz}$ as
$\ass{n}{\typn{\var0}}$ by \alles.

Next we define a function $T_0$ such that for natural numbers $n$ and
$m$, we have $T_0 \,\num 0\,\num m \betaetaeq \num m$, and $T_0 \,\num{n+1}
\,\num m \betaetaeq \num{m+1}$, as  
\[ T_0 := \abs{nxszs'z'}{\mathop{\chi_0}n(s(xsz)s'z')(xszs'z')} \]
The term $T_0$ can have the type $\natz \to \typn{\natz} \to \typn{\natz}$, 
which is verified as follows: in the context $\ass{x}{\typn{\natz}}$,
$\ass{s}{\natz\to\natz}$, $\ass{z}{\natz}$ we obtain the terms $xsz$ and
$s(xsz)$ of type $\natz$\@. These are instantiated with the rule \alles
as being of type $\typn{\var0}$, and with $\ass{s'}{\var0\to\var0}$ and
$\ass{z'}{\var0}$ we get $\ass{s(xsz)s'z'}{\var0}$ and $xszs'z':\var0$\@. 
Therefore we obtain 
\[ \judge{\Gamma}{\abs{s'z'}{\mathop{\chi_0} n (s(xsz)s'z')
    (xszs'z')}}{\typn{\var0}} \; , \] where $\Gamma$ is the context
$\ass{n}{\natz}, \ass{x}{\typn{\natz}}, \ass{s}{\natz\to\natz},
\ass{z}{\natz}$, and an application of \alli followed by several \impi
gives the claimed typing of $T_0$.  It is easily verified by
straightforward calculations that $T_0$ has the claimed operational
behaviour.

We use $T_0$ to implement an upward typecast that works with
the aid of a large parameter of suitable type, i.e., a term $\ass{\cu}
{\typnn{\natz} \to \natz \to \typn{\natz}}$ with the property that
$\cu \num m \, \num n \betaetaeq \num n$ as long as $m\geq n$\@. 
This can be implemented as 
\[ \cu := \abs{mn}{m\bigl(\abs{x}{T_0 (\sub n x) x}\bigr)\num0} \; , \]
i.e., the function $\abs{x}{T_0 (\sub n x) x}$, which operationally
behaves as 
\[ 
  \textsl{if $x < n$ then $x+1$ else $x$,}
\]
is iterated $m$ times, starting at $0$, to
the effect that in the first $n$ iterations, the value is increased by
$1$, and thereafter the value is $n$, and thus remains the same.

Now by use of the typecast, a more useful type-homogeneous
subtraction, but again with the aid of a large parameter, can be
defined as
 \[ \subt := \abs{mnk}{\cu m (\sub (\cd n) k)} \; : \;
\typnn{\natz} \to \typn{\natz} \to \typn{\natz} \to \typn{\natz} \; , 
\] 
with the property that $\subt \num m \, \num{n} \, \num{k} \betaetaeq
\num{n\monus k}$ as long as $m \geq n \monus  k$. 

\begin{definition}
  For a type $\tau$, let $\typi{\tau}{0} := \tau$, and
  $\typi{\tau}{k+1} := \typn{(\typi{\tau}{k})}$.
\end{definition}
To iterate the above construction, assume we have a subtraction 
\[ \ass{\subt_k }{ \typi{\natz}{k+1} \to \typi{\natz}{k} \to \typi{\natz}{k}
\to \typi{\natz}{k}} \; , \] 
and note that $T_0$ can have type $\typi{\natz}{k} \to
\typi{\natz}{k+1} \to \typi{\natz}{k+1}$ for every $k$, since $\chi_0$
can have type $\typi\natz{k+1}\to\typi\natz k\to\typi\natz k\to\typi\natz k$ (in
fact, $\chi_0$ can have any type of the form
$\typn\tau\to\tau\to\tau\to\tau$). 
Thus we can program an upward typecast 
\[ \cu_k := \abs{mn}{m \bigl(\abs{x}{T_0 (\subt_k (\cd m) n (\cd x)) x}
  \bigr)\num0}  \] 
of type ${\cu_k}\colon{\typi{\natz}{k+2} \to \typi{\natz}{k} \to
\typi{\natz}{k+1}}$, which again can be used to define a subtraction
\[ \subt_{k+1} :=  \abs{m{n_1}{n_2}}{\cu_k m \bigl( \subt_k (\cd m) (\cd
  n_1) (\cd n_2) \bigr)} \] 
of type $\typi{\natz}{k+2} \to \typi{\natz}{k+1} \to \typi{\natz}{k+1}
\to \typi{\natz}{k+1}$\@. 
Thus inductively we get subtractions $\subt_k$ and upward typecasts
$\cu_k$ for every $k$\@.
We also define iterated upward typecasts
$\ass{\cu^\ell_k}{ \typi{\natz}{k+\ell+1} \to \typi{\natz}{k} \to
\typi{\natz}{k+\ell}}$ by 
\[ \cu^0_k := \abs{mn}{n} \qquad \text{and} \qquad \cu^{\ell+1}_k :=
\abs{mn}{\cu_{k+\ell} m ( \cu^\ell_k (\cd m) n)} \; . \] 
From now on we will omit the index $k$ in $\cu_k$, $\cu^\ell_k$ and
$\subt_k$ when it can be inferred from the context.
We are ready to state our main lemma:
\begin{lemma}
  For every elementary recursive function $f\colon\Nat^n \to \Nat$ and
  $k\in \Nat$, there are a closed term $t$ and $\ell,r\in \Nat$ and a list
  $\vec{\eta}$ of types, where each $\eta$ is of the form $\eta ::=
  \typi{\natz}{k} \mid \typn{\eta} \mid \typn{(\eta \times \eta)}$,
  such that
  \[ \judge{}{t}{\vec{\eta} \to \vek{\typi{\natz}{k+\ell}} \to
    \typi{\natz}{k}} \] and for all $\vec{n} \in \Nat^n$, $t\,
  \num{\vec{L}} \, \num{\vec{n}} \; \betaetaeq \; \num{f(\vec{n})}$ as
  long as $L \geq 2_r(\sum \vec{n})$.
\end{lemma}
Note that we plug in the same numeral $\num L$ for all the arguments
of the types $\vec\eta$\@.  
Also note that only simple types over $\natz$ are used as these types
$\vec\eta$, and this is the only property used in the
application and proof. 
A statement similar to this lemma was offered by
Simmons~\cite{simmons} as a characterization of the Kalm\'ar
elementary recursive functions.

Before we prove the main lemma, we shall first use it to derive the
main theorem of this section, the representability of all elementary
recursive functions:
\begin{theorem} \label{elemdef}
  For every elementary recursive function $f\colon\Nat^n \to \Nat$ there is
  a closed term $\ass{T}{\nati^n \to \natz}$ such that $T\, \num{\vec{n}}
  \betaetaeq \num{f(\vec n)}$ for all $\vec n \in \Nat^n$. 
\end{theorem}
\begin{proof}
  From the lemma for $f$ and $k=1$, we obtain a term $t$ and
  $\ell,r,\vec{\eta}$ with the properties stated there. 
  As we can always move to bigger values of $r$, we may without loss of
  generality assume $r$ to be at least $2$ and even.
  
  Let $s := r/2$.  For each type $\eta_i$, instantiate each input
  $\ass{n}{\nati}$ as $\ass{n}{\typi{\eta_i}{s}}$, which is possible
  by \alled since $\typi{\eta_i}{s}$ is a closed type of level $1$\@.
  Now use $\ass{\add}{\typi{\eta_i}{s} \to \typi{\eta_i}{s} \to
    \typi{\eta_i}{s}}$ to compute $S := \sum \vec n$ of type
  $\typi{\eta_i}{s}$\@.  Next form the term $N :=
  (\ldots((S\num2)\num2)\ldots\num2)$, with $r$ occurrences of the
  numeral $\num2$, of type $\eta_i$.
  
  Instantiate the inputs $\vec n$ again by \alled at the closed, level $1$
  type $\typi{\natz}{\ell}$,  
  and form $T := \abs{\vec n}{\cd (t \, \vec{N} \, \vec{n})}$\@. 
  As for every input $\vec{n}$, $N$ evaluates to a numeral $\num L$ with
  $L \geq 2_r(\sum \vec n)$, the term $T$ has the required
  properties, by the lemma. 
\end{proof}
\begin{corollary}
  For every elementary recursive function $f\colon\Nat^n \to \Nat$ there is
  a closed term $\ass{T}{\nati^n \to \natz}$ such that $T\, \num{\vec{n}}
  \rd \num{f(\vec n)}$ for all $\vec n \in \Nat^n$. 
\end{corollary}
\begin{proof}
  Take $\abs{\vec nsz}T\vec nsz$ for the term $T$ obtained from the
  theorem. For every $\vec n$ consider the $\beta$-normal of
  $(\abs{\vec nsz}T\vec nsz)\num{\vec n}$, which is a closed,
  $\beta$-normal term of type $\Nat$, starting with two abstractions,
  hence a numeral. So it has to be $\num{f(\vec n)}$, for otherwise
  two distinct numeral would be $\beta\eta$-equal, which by the well
  known confluence of lambda calculus with pairs is not the case.
\end{proof}

\begin{proof}[Proof of the Lemma]
  We have produced terms representing the base functions successor,
  addition, subtraction and multiplication above.  For $S$, $+$ and
  $\times$, we can set $\ell=r=0$ and $\vec{\eta}$ empty for any
  $k$\@.  
  
  Concerning subtraction $\monus$, for $k=0$ we use the term $\abs{nk}{\sub
    (\cd n) k}$ and set $r=0$, $\ell=1$ and $\vec{\eta}$ empty, and
  for $k\geq 1$ we use $\subt_k$, and we set $\ell=r=0$ and
  $\vec{\eta}$ contains the single type $\typi{\natz}{k+1}$.
  
  In the following, note that by the properties of $\cd$, whenever we
  have a term $t$ of type $\typi{\natz}{k+\ell}$, we can obtain a term 
  $\cd^{\ell} t$ with the same value of type $\typi{\natz}k$. 

  For closure under composition, let $f(\vec n) = g(\vek{h_\cdot(\vec n)})$
  and $k$ be given. By the induction hypothesis for $g$ and $k$, we
  have a term $t_g$, numbers $\ell_g$ and $r_g$ and a list
  $\vek{\eta}$ of types such that the claim of the lemma holds for these. 
  
  Also, the induction hypothesis for each $h_i$ and $k+\ell_g$ yields
  terms $t_i$ and $\ell_i,r_i \in \Nat$ and types
  $\vek{\eta_{\cdot,i}}$, such that the claim holds for these. 
  
  Let $\ell := \ell_g + \max_i{\ell_i}$\@. Since the functions $h_i$
  are elementary recursive, $\sum h_i(\vec n)$ is also elementary,
  and therefore there is an $s\in\Nat$ such that $\sum_i
  {h_i(\vec n)} \leq
  2_s(\sum \vec n)$\@. For variables $\vec{v}$ and
  $\vek{w_{\cdot,i}}$, which we
  give the types $\vec\eta$ and $\vek{\eta_{\cdot,i}}$, respectively, 
   we set
  \[ t := \abs{\vec v \;\vec{\vec{w}}
    \;\vec n\,}{\, t_g \; \vec v
    \; \vek{(t_\cdot \:\vek{w_{\cdot,\cdot\cdot}}
   \: \vek{(\cd^{\ell-\ell_g-\ell_{\cdot\cdot}}\, n_\cdot)})}} \] 
   such that $t$ has type $\vec\eta \to \vec{\vec\eta} \to
  \vek{\typi{\natz}{k+\ell}} \to \typi{\natz}{k}$\@. By the induction
  hypothesis, for $r := \max(r_g + s,\vec{r})$ we have $t \,
  \vec{\num L} \, \vec{\vec{\num L}} \, \vec{\num n} \betaetaeq
  \num{f(\vec n)}$ as long as $L\geq 2_r(\sum \vec n)$.
  
  For closure under bounded sums, let $f(\vec n,m) = \sum_{i=0}^{m-1}
  g(\vec n,i)$ and $k$ be given. By the induction hypothesis for $g$
  and $k+1$, we have a term $t_g$, numbers $\ell$ and $r$ and a list
  $\vec{\eta}$ of types such that the claim of the lemma holds. 
  Define
  \[ \tilde{\chi}_0 := \abs{nxysz}{\chi_0\, n\, (xsz)\,(ysz)} \]
  of type $(\typi{\natz}{k+\ell+1})^3 \to \typi{\natz}{k+\ell+1}$,
  with the same operational semantics as $\chi_0$, i.e., for $i,j\in
  \Nat$ we have $\tilde{\chi}_0 \, \num0 \, \num i \,\num j \betaetaeq
  \num i$ and $\tilde{\chi}_0 \, \num{n+1} \, \num i \,\num j \betaetaeq
  \num j$\@.  For variables $\ass{v}{\typi{\natz}{k+\ell+2}}$, $\vec w$
  of the types $\vec\eta$ and $\vec n,m$ of type
  $\typi{\natz}{k+\ell+1}$, we have
\[ 
    T := \abs{x y}{\tilde{\chi}_0 \,
      (\subt v\, m\, y)x (\add x (\cu^\ell \, v \, (t_g \,\vec w 
      \,\vec n \,y)))} \; , 
\]
of type
$\typi{\natz}{k+\ell+1}\to\typi{\natz}{k+\ell+1}\to\typi{\natz}{k+\ell+1}$.

As long as a sufficiently large numeral $\num L$ is substituted for
the variables $v$ and $\vec w$, $T$ operationally behaves as 
\begin{quote}
  \textsl{if $y< m$ then $x + g(\vec n, y)$ else $x$}.
\end{quote}
More precisely, $L$ has to be large enough so that all values of
$g(\vec n,i)$ are computed correctly, that is, $L \geq 2_r(\sum\vec n
+ m)$, and we need $L \geq g(\vec n,i)$ for the typecast $\cu^\ell$ to
work properly.  Next, we define
 \[  P := \abs{p}{\pair{T (\pl{p}) \,(\pr{p})}{\suc (\pr{p})}}  \] 
 of type $(\typi{\natz}{k+\ell+1} \times \typi{\natz}{k+\ell+1}) \to
 (\typi{\natz}{k+\ell+1} \times \typi{\natz}{k+\ell+1})$\@.  When this
 term, having the operational semantics
\[ \pair{s}{i} \mapsto  \begin{cases} \pair{s+g(\vec n,i)}{i+1}
  &\text{if } i<m \\ \pair{s}{i+1} &\text{otherwise,} \end{cases} \]
is iterated starting from the pair $\pair{0}{0}$, by use of a
sufficiently large numeral of type $\typn{(\typi{\natz}{k+\ell+1}
  \times \typi{\natz}{k+\ell+1})} $, the values $g(\vec n,i)$ for
$i=0,\ldots,m-1$ are summed up in the left component.  Thus to
represent $f$, we define the term
\[
    t := \abs{u\, v \, \vec w \, \vec n \, m}
             { \cd^{\ell+1} \bigl(\pl{
                  u \, P \, \pair{\num0}{\num0}
                      }\bigr)} 
\]
of type
\[ 
    \typn{(\typi{\natz}{k+\ell+1} \times \typi{\natz}{k+\ell+1})} \to
    \typi{\natz}{k+\ell+2} \to \vec\eta \to \vek{\typi{\nat}{k+\ell+1}} \to
    \typi{\natz}{k} 
\]
By the induction hypothesis and the construction, we get the property
that $t \, \num L \, \num L \, \vec{\num L} \, \vec{\num n} \, \num m
\; \betaetaeq \num{f(\vec n , m )}$ as long as $L$ is sufficiently large.
To be more precise, $L$ needs to satisfy the requirements above for $T$ to be
computed correctly, and $L \geq m$ in order to complete the summation\@.  
Therefore, let $s$ be such that for every $m$ and $i\leq m$ we have
$g(\vec n,i) \leq 2_s(\sum\vec n+m)$, which exists since $g$ is
elementary recursive, and let $r' := \max(r,s)$\@. Then all
conditions on $L$ are satisfied if $L \geq 2_{r'}(\sum \vec n + m)$.

Closure under bounded products is shown in the same way, only with
$\add$ in the definition of $T$ replaced by $\mul$, and the iteration
of $P$ is started at $\pair{1}{0}$.  
\end{proof}
\end{section}

\begin{section}{Soundness}

In this section we show the other direction of our claim, that is, we
show that every term of type $\nati\to\natz$ denotes a function on
Church numerals computable in elementary space.  The main idea is to
use the elementary bound for traditional cut-elimination in
propositional logic. In this section we will deal only with types of
level at most $1$, so let $\tau$, $\rho$, $\sigma$ range over those
types within this section. Note that every instantiation of $\nati$ is
a type of level $1$\@. 
Types of level $0$ and $1$ are almost simple types
(corresponding to propositional logic) with the exception of
quantification of $\var0$\@. These quantifiers however, can only be
instantiated with flat types of the form $\var0\times\ldots\times\var0$\@.
Hence we can get a notion of cut-rank that is invariant under
generalization and instantiation of level $0$, if we ignore pairs.
Fortunately we can do so, as the reduction of a pair-redex reduces the
size of the term and hence does not do any harm. So we define the rank
$\rk{\tau}$ of a type $\tau$ inductively as follows:
\begin{align*}
  \rk{\alpha} &:= 0 \\
  \rk{\rho\times\sigma} &:= \max(\rk{\rho},\rk{\sigma}) \\
  \rk{\rho\to\sigma} &:= \max(\rk{\rho}+1,\rk{\sigma}) \\
  \rk{\all{\alpha}{\rho}} &:= \rk{\rho} 
\end{align*}
We inductively define a relation $\ajudge{\Gamma}{m}{k}{r}{\tau}$
saying that $\judge{\Gamma}{r}{\tau}$ can be derived by a typing
derivation of height $m$ and cut-rank $k$:
\begin{align*}
  &\ax\:\frac{}{\ajudge{\Gamma}{m}{k}{x}{\tau}} & &\text{if }x:\tau\text{ occurs in } \Gamma \text{ and } m,k \geq 0  
\\[2ex] 
  &\impi\:\dfrac{\ajudge{\Gamma,x:\sigma}{m}{k}{r}{\rho}}{\ajudge{\Gamma}{m+1}{k}{\abs x r}{\sigma\to\rho}}  
\\[2ex] 
  &\impe\:\dfrac{\ajudge{\Gamma}{m}{k}{r}{\sigma \to \rho} \qquad
    \ajudge{\Gamma}{m'}{k}{s}{\sigma}}{\ajudge{\Gamma}{m''}{k}{rs}{\rho}} 
  & &\text{if } \rk{\sigma}<k 
\\[2ex]
  &\prodi\:\dfrac{\ajudge{\Gamma}mk{r}{\rho} \qquad \ajudge\Gamma{m'}ks\sigma}{\ajudge\Gamma{m''}k{\pair{r}{s}}{\rho\times\sigma}} 
\\[2ex] 
  &\prodel\:\dfrac{\ajudge\Gamma mkr{\sigma\times\rho}}{\ajudge\Gamma{m+1}k{\pl r}\sigma} & &\text{and analogous for \proder} 
\\[2ex]
  &\alli\:\dfrac{\ajudge{\Gamma}mkr\tau}{\ajudge{\Gamma}{m+1}kr{\all{\alpha}{\tau}}} 
  & &\text{if } \alpha \notin \fv{\Gamma}
\\[2ex]  
  &\alles\:\dfrac{\ajudge\Gamma mkr{\all{\alpha}{\tau}}}
  {\ajudge\Gamma{m+1}kr{\subst{\tau}{\alpha}{\sigma'}}}
  & &\text{where } \sigma' \text{ is a flat type.}
 \end{align*}
 where $m'' := \max(m,m')+1$\@. As the rules are precisely those of
 our typing judgment for types of level at most $1$, we have the
 following property for typing derivations of level at most $1$: if
 $\judge\Gamma r\tau$ then there are $m$, $k$ such that $\ajudge\Gamma
 mkr\tau$\@.  On the other hand, the following property obviously
 holds and motivates our interest in this notion:
\[ \ajudge\Gamma mkr\tau \quad \text{implies} \quad |r| \leq 2^m \]
The rules are formulated in such a way that weakening is admissible.
\begin{proposition}[Weakening]
If $\ajudge\Gamma mkr\tau$, $\Gamma'\supset\Gamma$, $m'\geq m$, $k'\geq
k$ then $\ajudge{\Gamma'}{m'}{k'}r\tau$.
\end{proposition}
The next proposition, which can be shown by a trivial induction on
$\typp$ or $\tau$, respectively, explains formally why we can allow
instantiations with flat types of level $0$ without any harm: the rank
is not altered!
\begin{proposition}\label{prop-rank-subst} For a flat type $\typp$ of
  level $0$ we have $\rk{\typp}=0$ and
  $\rk{\subst\tau{\var0}{\typp}}=\rk\tau$.
\end{proposition}
Knowing that the rank of a type is not altered by substituting in a
flat type, the cut-rank, being a rank, is not altered as
well, hence an induction on $\ajudge\Gamma mkt\tau$ shows:
\begin{proposition}\label{prop-deriv-subst}
If $\ajudge\Gamma mkt\tau$ and $\typp$ is a flat type of level $0$ then
$\ajudge{\subst\Gamma{\var0}{\typp}}mkt{\subst\tau{\var0}{\typp}}$
\end{proposition}
Using this proposition a simple induction on $m$ shows that a
derivation $\ajudge\Gamma mkt\tau$ can be transformed in such a way
that the rule $\alli$ is never followed by $\alles$\@.
So from now on we tacitly assume all derivations to be free from
those $\alli$-$\alles$-redexes, as for example in the
proof of the next proposition, which then is a simple analysis of the
last rule of the derivation.
\begin{proposition}
If $\ajudge\Gamma mk{\pl{\pair rs}}\rho$ then $\ajudge\Gamma mkr\rho$ and
if $\ajudge\Gamma mk{\pr{\pair rs}}\sigma$ then $\ajudge\Gamma
mks\sigma$.
\end{proposition}
As usual, induction on the first derivation shows that cuts can be
performed at the cost of summing up heights.
\begin{lemma}\label{lem-cut}
If $\ajudge{\Gamma,\ass x\rho}mks\sigma$ and 
$\ajudge\Gamma{m'}kr\rho$ then $\ajudge\Gamma{m+m'}k{\subst sxr}\sigma$.
\end{lemma}
In order to be able to reduce the cut rank, we first show an
``inversion''-lemma, that is, we show that under certain conditions
terms of arrow-type can be brought into abstraction form.
\begin{lemma}[Inversion]
If $\rk\Gamma\leq k$ and $\ajudge\Gamma mkt{\rho\to\sigma}$ where
$\rk\rho\geq k$, then there are $t'$ and $x$ with $t\betaeq\abs x{t'}$
such that $\ajudge{\Gamma,\ass x\rho}mk{t'}\sigma$.
\end{lemma}
\begin{proof}
Induction on $m$ and case distinction according to $t$.

The case $t=x\vec s$ is impossible, since $x$ would have to occur in
$\Gamma$ and hence $\rk\Gamma>k$\@.  The case $t=\pair rs\vec t$ is
also impossible since $\vec t$ has to be empty, as we assume $t$ to be
free of pair-redexes, and therefore $t$ would have to have a pair type.

So the only remaining case is that $t$ is of the form $t=(\abs yr)\vec
t$\@. The claim is trivial if $\vec t$ is empty. So without loss of
generality we might assume $t$ to be $t=(\abs y r)s\vec s$, with
$y$ not free in $s,\vec s$\@. The abstraction $\abs yr$ must
have been introduced from a derivation
$\ajudge{\Gamma,\ass y\tau}mkr{\tilde\tau}$ with $\rk\tau<k$ for otherwise
the cut would not have been allowed. Hence, for some $m'$ with
$m'+2\leq m$ we get 
\[
\ajudge{\Gamma,\ass y\tau}{m'}k{r\vec
  s}{\rho\to\sigma} \text{~and~} \ajudge\Gamma{m'+1}ks\tau
\]
Hence by the
induction hypothesis we get a new variable $x$ and a term $t'$ such
that $r\vec s\betaeq\abs x{t'}$ and
$\ajudge{\Gamma,\ass y\tau,\ass x\rho}{m'}k{t'}\sigma$\@. From that we
conclude $\ajudge{\Gamma,\ass x\rho}{m'+2}k{(\abs y{t'})s}\sigma$ and
note $\abs x{(\abs y{t'})}s\betaeq\abs x{\subst{t'}ys}=\subst{(\abs
  x{t'})}ys \betaeq\subst{(r\vec s)}ys=\subst rys\vec s\betaeq(\abs
yr)s\vec s=t$, hence the claim.
\end{proof}

\begin{lemma}[Cut-rank reduction]
If $\ajudge\Gamma m{k+1}t\rho$, $\rk\Gamma\leq k$, and $\rk\rho\leq
k+1$ then $\ajudge\Gamma{2^m}k{t'}\rho$ for some $t'\betaeq t$.
\end{lemma}
\begin{proof}
Induction on $m$\@. The only interesting cases are $\impi$ and
$\impe$\@. Concerning $\impi$ we are in the situation that
$\ajudge\Gamma{m+1}{k+1}{\abs xr}{\sigma\to\tau}$ was concluded from
$\ajudge{\Gamma,\ass x\sigma}m{k+1}r\tau$\@. With $\rho=\sigma\to\tau$ we
have $\rk\Gamma\leq k$, $\rk\sigma<\rk\rho\leq k+1$ and
$\rk\tau\leq\rk\rho\leq k+1$\@. Hence an application of the induction
hypothesis yields $\ajudge{\Gamma,\ass x\sigma}{2^m}k{r'}\tau$ from
which we conclude $\ajudge{\Gamma}{2^m+1}k{\abs x{r'}}{\sigma\to\tau}$
which, by weakening, suffices, since $2^m+1\leq2^{m+1}$.

Concerning the case $\impe$ we are in the situation that
$\ajudge\Gamma{m+1}{k+1}{ts}\rho$ was concluded from $\ajudge\Gamma
m{k+1}t{\sigma\to\rho}$ and $\ajudge\Gamma m{k+1}s\sigma$\@. The only
case that is not immediate by the induction hypothesis is if
$\rk\sigma=k$\@. Then the induction hypothesis gives us
$\ajudge\Gamma{2^m}k{t'}{\sigma\to\rho}$ for some $t'\betaeq t$\@. By
our assumption $\rk\Gamma\leq k$, hence by inversion we get
$\ajudge{\Gamma,\ass x\sigma}{2^m}k{t''}\rho$ for some new $x$ and $t''$
such that $\abs x{t''}\betaeq t'\betaeq t$\@. Also by the induction
hypothesis we get $\ajudge\Gamma{2^m}ks\sigma$\@. By Lemma~\ref{lem-cut}
we get $\ajudge\Gamma{2^m+2^m}k{\subst{t''}xs}\rho$ which yields the
claim since $\subst{t''}xs\betaeq(\abs x{t''})s\betaeq ts$.
\end{proof}

\begin{corollary}\label{cor-cut-elim}
If $\ajudge{}m{k+1}t{\typn{\var{}}}$ then
$\ajudge{}{2_k(m)}1{t'}{\typn{\var{}}}$ for some $t'\betaeq t$\@.
\end{corollary}

\begin{proposition}\label{prop-lamfree}
If $t$ normal and $\judge\Gamma t{\typp}$ for some $\Gamma$ with
$\rk\Gamma\leq1$ then $t$ is $\lambda$-free.
\end{proposition}
\begin{proof}
Inspection of the typing rules yields that the only rule introducing a
$\lambda$ is $\impi$, which creates an arrow-type. In order for the
whole term to be of arrow-free type, the rule $\impe$ has to be used,
either creating a redex or requiring a variable of rank at least $2$.
\end{proof}

\begin{definition}
A term $t$ is \emph{quasinormal}, if every redex in $t$ is of the form
$\pl{\pair rs}$ or $\pr{\pair rs}$ with $\lambda$-free $r$ and $s$
\end{definition}
We remark the trivial property that the normal form of a quasinormal
term $t$ can be computed in space bound by the length of $t$\@. We also
note that Proposition~\ref{prop-lamfree} also holds for quasinormal
terms, since the only types discarded by a redex are those of terms
which are $\lambda$-free by definition. Moreover, a simple induction
on $t$ shows:
\begin{proposition}
If $t$ is quasinormal and $s$ is $\lambda$-free and quasinormal then
$\subst txs$ is quasinormal.
\end{proposition}
From that proposition, Proposition~\ref{prop-lamfree} and
Lemma~\ref{lem-cut} we immediately get:
\begin{corollary}\label{cor-qn-cut}
If $\ajudge{\Gamma,\ass x\sigma'}m1r\rho$ and
$\ajudge\Gamma{m'}1s{\sigma'}$ and $r$ and $s$ are quasinormal then
$\ajudge\Gamma{m+m'}1{\subst rxs}\rho$ and $\subst rxs$ is
quasinormal.
\end{corollary}
This corollary allows us to show our last ingredient for the soundness
theorem: we can transform a term with cut-rank $1$ into a quasinormal
one at exponential cost.
\begin{lemma}\label{lem-qn-cut-elim}
If $\ajudge\Gamma m1t\tau$ then $\ajudge\Gamma{2^m}1{t'}\tau$
for some quasinormal $t'$ with $t'\betaeq t$.
\end{lemma}
\begin{proof}
Induction on $m$\@. If $t$ is not quasinormal, it has a subterm of the
form $(\abs xr)s$\@. Then, for some $\Delta$, $\sigma$, $\rho$ and $k$ we have
$\ajudge{\Delta,\ass x\sigma}k1r\rho$, and $\ajudge\Delta{k+1}1s\sigma$
from which $\ajudge\Delta{k+2}1{(\abs xr)s}\rho$ was concluded. Since
the cut was allowed, we have $\rk\sigma<1$\@. Hence, by the induction
hypotheses we get a quasinormal
$s'\betaeq s$ such that $\ajudge\Delta{2^{k+1}}1{s'}\sigma$\@. Also by
induction hypothesis we get a quasinormal $r'\betaeq r$ such that
$\ajudge{\Delta,\ass x\sigma}{2^k}1{r'}\rho$\@. By
Corollary~\ref{cor-qn-cut} we get
$\ajudge\Delta{2^{k}+2^{k+1}}1{\subst{r'}x{s'}}\rho$ and
${\subst{r'}x{s'}}$ is quasinormal, hence the claim.
\end{proof}
We are now ready to show that every representable function is
elementary recursive. To keep the notation simple, we only state and
prove this for unary functions, but the generalization to higher
arities is straightforward.
\begin{theorem}
If $\judge{}t{\nati\to\natz}$ then $t$ denotes an elementary function
on Church numerals.
\end{theorem}
\begin{proof}
We have $\judge{\ass x\nati}{tx}\natz$\@. Since all our terms are also
typeable in usual system $F$, hence strongly normalizing, and since
subject reduction holds, we can find (in maybe long time, which
however is independent of the input) a normal term $t'\betaeq tx$ and
$\judge{x:\nati}{t'}\natz$\@.
Since $t'$ is normal, inspection of the typing rules yields that every
occurrence of $x$ must be within some context, that is, of the form 
$$
\alled\quad
\frac{\judge{\ass x\nati}x\nati}{\judge{\ass x\nati}x{\typn\xi}}
$$ 
for some level $1$ type $\xi$, without (free) variable $\var1$\@. Let
$c$ be the maximum of the ranks of all the $\xi$'s occurring in that
derivation and $k$ the number of occurrences of such $\xi$'s (note that
$c$ and $k$ are still independent of the input).

Now, let a natural number $n$ be given. Replacing all $\ass x{\typn\xi}$
by derivations of $\ass{\num n}{\typn\xi}$ yields a term $t''\betaeq
t\num n$ and a derivation
$\ajudge{}{k\cdot(n+2)+2|t'|}c{t''}\natz$\@. The bound on the height of
the derivation is obtained as follows: 
there are $k$ derivations of height $n+2$ yielding $\ass n{\typn\xi}$
and these are plugged into the derivation of $\ass{t'}\natz$\@. In the
latter derivation there is at most one inference for each symbol in
$t'$ followed possible by a single quantifier inference.

Using Corollary~\ref{cor-cut-elim} we obtain a term $\tilde t\betaeq
t''\betaeq t\num n$ such that\linebreak
$\ajudge{}{2_{c+1}(k(n+2)+2|t'|)}1{\tilde t}\natz$\@. Hence
Lemma~\ref{lem-qn-cut-elim} and the remark on computing the normal
form of a quasinormal term provides means to calculate the normal form
of $t\num n$ in elementary space. (Note that all the intermediate terms
are also of elementary bounded size.)
\end{proof}

Together with Theorem \ref{elemdef} we obtain the claimed characterization.
\begin{corollary} \label{char}
  The representable functions are precisely the elementary recursive
  functions. 
\end{corollary}

Note that our characterization does not mean that the normalization
procedure for terms typeable in our system is elementary recursive.
The following easy counterexample shows that this is indeed not the
case: the terms $(\ldots ((\num 2 \,\num 2)\num 2) \ldots \num 2 )$ with $n$
occurrences of $\num 2$ are of size $O(n)$, but their normal forms are
the numerals $\num{2_n(1)}$ of size $\Omega(2_n(1))$\@. Thus the
normalization function has super-elementary growth. 
\end{section}


\begin{thebibliography}{10}

\bibitem{altcoq01}
T.~Altenkirch and T.~Coquand.
\newblock A finitary subsystem of the polymorphic lambda-calculus.
\newblock In S.~Abramsky, editor, {\em Typed Lambda Calculi and Applications},
  volume 2044 of {\em LNCS}, pages 22--28. Springer, 2001.

\bibitem{asprov02}
A.~Asperti and L.~Roversi.
\newblock Intuitionistic light affine logic.
\newblock {\em ACM Transactions on Computational Logic}, 3(1):137--175, 2002.

\bibitem{clotehb}
P.~Clote.
\newblock Computation models and function algebras.
\newblock In E.~R. Griffor, editor, {\em Handbook of Computability Theory},
  pages 589--681. Elsevier, 1999.

\bibitem{danjoi02}
V.~Danos and J.-B. Joinet.
\newblock Linear logic and elementary time.
\newblock {\em Information and Computation}, 183(1):123--137, 2003.

\bibitem{gir71}
J.-Y. Girard.
\newblock Une extension de l'interpr\'etation de {G\"o}del \`a l'analyse, et
  son application \`a l'\'elimination des coupures dans l'analyse et la
  th\'eorie des types.
\newblock In J.~Fenstad, editor, {\em Proceedings of the 2nd Scandinavian Logic
  Symposium}, pages 63--92. North-Holland, 1971.

\bibitem{gir98}
J.-Y. Girard.
\newblock Light linear logic.
\newblock {\em Information and Computation}, 143(2):175--204, 1998.

\bibitem{grz53}
A.~Grzegorczyk.
\newblock Some classes of recursive fuctions.
\newblock {\em Rozprawy Matematyczne}, 4, 1953.

\bibitem{kalmar}
L.~Kalm\'ar.
\newblock Egyszer{\H{u}} p\'elda eld{\"o}nthetetlen aritmetikai probl\'em\'ara.
\newblock {\em Matematikai \'es Fizikai Lapok}, 50:1--23, 1943.

\bibitem{leiv91}
D.~Leivant.
\newblock Finitely stratified polymorphism.
\newblock {\em Information and Computation}, 93:93--113, 1991.

\bibitem{rey74}
J.~Reynolds.
\newblock Towards a theory of type structure.
\newblock In {\em Proceedings, Colloque sur la programmation}, pages 408--425.
  Springer LNCS 19, 1974.

\bibitem{simmons}
H.~Simmons.
\newblock Tiering as a recursion technique.
\newblock \emph{Bulletin of Symbolic Logic}, to appear.

\end{thebibliography}
\end{document}